\documentstyle[epsfig,12pt,epsf,cite]{article}

\newcommand{\delslash}{\partial \hspace{-6pt}/}
\newcommand{\mevsig}{\langle \sigma \rangle}
\begin{document}

\begin{center}
{\large\bf
$\eta$-Nucleus interactions
and in-medium properties of N$^*$(1535)
in chiral models
}

\vspace{3mm}

{H.~Nagahiro$^a$, D.~Jido$^b$\footnote{
Present address: European Centre for Theoretical Studies in Nuclear
Physics and Related Areas (ECT*), Villa Tambosi, Strada delle Tabarelle
286, I-38050 Villazzano (Trento), Italy
} and S.~Hirenzaki$^a$}
\vspace{3mm}

{\small \em
$^a$Department of Physics, Nara Women's University, Nara 630-8506, Japan\\
$^b$Research Center for Nuclear Physics, Osaka University, \\
Ibaraki, Osaka 567-0047, Japan}
\end{center}

\abstract{
The properties of $\eta$-nucleus interaction and their experimental
consequences  are investigated
with $\eta$-nucleus optical potentials obtained by postulating the 
$N^*(1535)$ dominance for $\eta$-$N$ system. The $N^*(1535)$ properties
in the nuclear medium are evaluated by two kinds of chiral effective models
based on distinct pictures of $N^*(1535)$. We find that these two models 
provide qualitatively different optical potentials of the $\eta$ meson,
reflecting the in-medium properties of $N^*(1535)$ in these
models. In order to compare these models in physical observables, we 
calculate spectra of (d,$^3$He) reactions for the $\eta$ mesic nucleus
formation with various kinds of target nuclei.  We show that the
(d,$^3$He)
spectra obtained in these models are significantly different
and are expected to be distinguishable in experiments.


}

\section{Introduction}

The study of the in-medium hadron properties is one of the most interesting
subjects in nuclear physics and has attracted continuous attention for
decades.
Historically, several kinds of hadron-nucleus bound systems were
investigated such as
pionic atoms, kaonic atoms, and $\bar{p}$ atoms \cite{ref:PR287}.
Recently, the interests and importance of this field have been much increased
due to both the theoretical and experimental developments.

One of the important progresses in theoretical aspects is the new concept of
partial restoration of chiral symmetry \cite{ref:PR247},
in which a reduction of the order parameter of the chiral phase transition
takes place in finite density and causes the modifications of the hadron properties. 
The development of the chiral effective theories enables us to discuss
in-medium properties of hadrons in the  viewpoint of chiral symmetry.
In this context,
hadronic bound systems have been investigated in various chiral models for
pionic atoms \cite{ref:PLB514,ref:APP31}, kaonic atoms/nuclei \cite{ref:PRC61},
$\eta$ and $\omega$ mesic nuclei \cite{ref:PRC66,ref:PLB550,ref:NPA650}.

Experimentally, the establishment of the (d,$^3$He) spectroscopies in
the formation of
the deeply bound pionic atom opens new possibilities to form
various kinds of hadron-nucleus bound systems, which are not accessible
by the standard X-ray spectroscopies, and to investigate the bound
states quite precisely.
Originally the (d,$^{3}$He) reaction has been studied 
theoretically \cite{ref:NPA1991} as one of the proper methods
to form the deeply bound pionic states \cite{ref:PL1988}, and later proved to be a
powerful tool experimentally \cite{ref:K.Itahashi}.
Using the (d,$^3$He)
reactions the deeply bound pionic 1s states were observed clearly and
the binding energies and widths are determined precisely
\cite{ref:PLB514,ref:K.Itahashi,ref:Geiseel}.  This method can be
generalized to
form other hadron-nucleus bound systems
\cite{ref:PRC66,ref:EPJA6,ref:PLB443}.
An experimental proposal to observe the $\eta$ and $\omega$-nucleus
system by the 
(d,$^3$He) reactions at GSI has been approved already
\cite{ref:GSI}.



We investigate the $\eta$ mesic nucleus in this paper.
The special features of the $\eta$ mesic nucleus are the followings:
(1) the $\eta-N$ system dominantly couples to $N^{*}(1535)$ 
$(N^{*})$ at the threshold
region \cite{ref:PRD66}. (2) The isoscalar particle $\eta$ filters out
contamination of the isospin$-3/2$ excitation in the nuclear medium. (3)
As a result of the $s$-wave nature of the $\eta NN^*$ coupling, there is no
threshold suppression like the $p$-wave coupling. 

The dominant coupling of $\eta N$ to $N^*(1535)$ 
makes the use of this channel particularly suited to investigate this resonance and 
enables us to consider the $\eta$ mesic nucleus as one of the doorways to 
investigate the in-medium properties of the $N^*$.
As shown in Ref. \cite{ref:PRC66}, the $\eta$-nucleus optical
potential is extremely sensitive to the in-medium masses of $N$ and
$N^*$ and even its qualitative nature may change from attractive to repulsive.

In this paper, we calculate the $\eta$ nucleus optical potential
assuming the $N^{*}$ dominance in $\eta$-$N$ system, and use the chiral doublet
models (the naive and mirror models) \cite{ref:PRD39,ref:PTP106}
and the chiral unitary model \cite{ref:PLB550}
to calculate the in-medium modification of $N^{*}$.
These models are based on quite different pictures of $N^{*}$.
In the chiral doublet model, 
the $N$ and $N^*$ form a multiplet of the
chiral group. In Refs. \cite{ref:PLB224, ref:NPA640}, a reduction of the
mass difference of the $N$ and $N^*$ in the nuclear medium is found in
the chiral doublet model. On the other hand, an investigation of the
$\eta$ meson properties in the nuclear medium within a chiral unitary
approach has been also reported \cite{ref:PLB550}. 
There the $N^*$ is introduced as a resonance generated dynamically
from meson-baryon scattering. Since this
theoretical framework is quite different from the chiral doublet model, 
it is interesting to compare the
consequences of these `chiral' models for $N^*$ and $\eta$ mesic
nucleus. 

For this purpose we calculate
the (d,$^3$He) spectra for various cases and show the numerical results. 
We find the significant differences for the spectra and can
expect to distinguish the models from the experimental observables.
Since the optical potential for the $\eta$ predicted with the chiral doublet model
may change its nature from attractive to repulsive for higher nuclear
densities, we even consider the $\eta$ bound systems for
unstable nuclei which are known to have low density halo structure in some
nuclides.
We would like to emphasize that we have the possibilities to deduce the
$\eta$-nucleus optical potential information from the experiments and
obtain the $N^*$ property in medium which has the close connection to
the $N$-$N^*$ chiral dynamics.

In section \ref{sec:model}, we describe the $\eta$-nucleus optical potentials 
obtained in the chiral doublet model with the naive and mirror assignments
and in the chiral unitary model with assuming the $N^*$ dominance in
$\eta N$ channel. In section 3, we show the calculated (d,$^3$He)
spectra for the formation of the $\eta$-nucleus systems. Section 4 is
devoted to the summary.

\section{Chiral models for $\eta$-nucleus Interaction}
\label{sec:model}
%

In this section, we show the formulation to calculate the $\eta$ optical
potential in a nucleus. We use the chiral models that incorporate chiral 
symmetry in different way in order to evaluate the in-medium behaviors
of the $N^{*}$ resonance. 

In the recoilless (d,$^{3}$He) reaction, which is proton picking-up
process, the $\eta$ meson can be created in the nucleus with small momentum.
Therefore here we assume the $\eta$ meson at rest in the nucleus.

\subsection{General features of $\eta$-nucleus optical potential}
First of all, we consider the $\eta$-nucleus optical potential
within the $N^*$ dominance hypothesis in the $\eta$-nucleon channel
as discussed in Sec.1.
If we assume the Lagrangian formulation for $N^*$, where $N^*$ is
described
as a well-defined field and its propagator is written in the Breit-Wigner
form, it is shown as a general conclusion that the
$\eta$-nucleus optical potential is very sensitive to the in-medium
difference of the $N$ and $N^*$ masses.

Consider an analogy to the
isobar model for the $\Delta$ resonance in $\pi$-$N$ system,
we obtain the $\eta$-optical
potential in the nuclear medium in the heavy baryon limit \cite{ref:PRC44}
as;
\begin{equation}
V_\eta(\omega)=\frac{g_\eta^2}{2\mu}\frac{\rho(r)}
{\omega+m^*_{N}(\rho)-m^*_{N^*}(\rho)+i\Gamma_{N^*}(s;\rho)/2},
\label{eq:potential}
\end{equation}
where $\omega$ denotes the $\eta$ energy and $\mu$ is the reduced mass
of the $\eta$ and the nucleus. $\rho(r)$ is the density distribution
for nucleus. 
The $\eta NN^{*}$ vertex with the coupling constant $g_{\eta}$ is given by
\begin{equation}
L_{\eta NN^*}(x)=g_\eta \bar{N}(x)\eta(x)N^*(x) + {\rm h.c.},
\end{equation}
where $g_{\eta} \simeq 2.0$ to reproduce the partial width $\Gamma_{N^*
\rightarrow \eta N} \simeq 75$ MeV\cite{ref:PRD66} at tree level.
The ``effective masses ''$m^*_N(\rho)$ and $m^*_{N^*}(\rho)$ 
in the medium are 
defined as poles of their
propagators so that ${\rm Re\ }G^{-1}(p^0=m^*,\vec{p}=0)=0$. 
Considering 
that the $N^*$ mass in free space lies only fifty
MeV above the $\eta N$ threshold and
that in the medium the mass difference of
$N$ and $N^*$ becomes smaller in the chiral double model, 
we expect that there is a critical
density $\rho_c$ where the sign of $\omega + m^*_N - m^*_{N^*}$ is
getting to be positive. Then, the $\eta$-nucleus optical potential turns
to be repulsive at density above the $\rho_c$.

\subsection{Chiral doublet model}
\label{subsec:CDM}
In this subsection, we evaluate the effective masses of $N$ and $N^*$ using
the chiral doublet model with the mirror and naive assignments in order to
obtain the $\eta$-nucleus optical potential.

The chiral doublet model is an extension of the SU(2) linear sigma model
for the nucleon sector. There are two possible models in the chiral doublet model;
the naive and mirror models \cite{ref:PTP106, Nemoto:1998um}. In the later model, $N^*$ is regarded as 
chiral partner of $N$ and forms a chiral multiplet together with $N$.
The Lagrangian density of the chiral doublet model with the mirror
assignment is given by
\begin{eqnarray}
  {\cal L} &=& \sum_{j=1,2} \left[ \bar{N}_j i \delslash N_j -
  g_j \bar{N}_j ( \sigma + (-)^{j-1}i\gamma_5 \vec \tau \cdot \vec \pi) N_j
  \right] \nonumber \\
  && - m_0 (\bar N_1 \gamma_5 N_2 - \bar N_2 \gamma_5 N_1)
  \label{eq:mirmodel}
\end{eqnarray}
where $N_1$ and $N_2$ are eigenvectors under the SU(2) chiral
transformation.
This Lagrangian was proposed and investigated first by
DeTar and Kunihiro \cite{ref:PRD39}.
In the mirror assignment, $N_1$ and $N_2$ have an opposite
axial charge to each other while they have the same charge in the `naive'
assignment, 
which is discussed later. 
The physical $N$ and $N^*$ are expressed as a
superposition of $N_1$ and $N_2$ as $N=\cos\theta N_1+\gamma_5 \sin\theta
N_2$ and $N^*=-\gamma_5 \sin\theta N_1 + \cos\theta N_2$ where $\tan
2\theta=2m_0/\mevsig(g_1+g_2)$ \cite{ref:PTP106},
in order to diagonalize the mass terms after spontaneous breaking of
chiral symmetry. The $N$ and $N^*$ masses
are given by
\begin{equation}
    m^{*}_{ N,N^*}
    = {1 \over 2} ( \sqrt{ (g_1 + g_2)^2 \mevsig^2 +
    4 m_0^2 } \mp (g_2-g_1) \mevsig ) \label{eq:mass},
\end{equation}
and the coupling constant of the $\pi NN^*$ vertex also is given by
\begin{equation}
    g_{\pi NN^*} = (g_2-g_1)/\sqrt{4 + ((g_1 + g_2)\mevsig/m_0)^2},
    \label{eq:coup}
\end{equation}
where $\mevsig$ is the sigma condensate in the nuclear medium.
The parameters in the Lagrangian have been chosen so that the
observables in vacuum, $m_N=940$ MeV, $m_{N^*}=1535$ MeV, and
$\Gamma_{N^*\rightarrow \pi N}\simeq 75$ MeV, are reproduced with
$\mevsig_0=f_\pi=93$ MeV, and they are obtained as $g_1=9.8$,
$g_2=16.2$, $m_0=270$ 
MeV \cite{ref:PTP106}. 
It is important here that the masses and couplings of $N$ and $N^*$ are
constrained by chiral symmetry and are written as functions of the sigma
condensate. Such constraints are also observed in the chiral quartet
model for $\Delta(1232)$ and $N(1520)$ with $J=3/2$ \cite{Jido:1999hd}.

Assuming partial restoration of chiral symmetry in the nuclear medium,
we parameterize the sigma condensate as a function of the nuclear
density $\rho$ 
as 
\begin{equation}
\mevsig = \Phi(\rho)\mevsig_0,
\label{eq:sigmainmedium}
\end{equation}
where, in the linear density approximation, $\Phi(\rho)=1-C\rho/\rho_0$
with $C=0.1\sim0.3$ \cite{ref:PRL82}. The $C$ parameter represents the
strength of the chiral restoration at the nuclear saturation density
$\rho_0$.
In the mean field approximation, the medium effects may be introduced
by replacing the in-vacuum sigma condensate by the in-medium one.
Finally the in-medium mass difference is obtained by
\begin{equation}
m^*_N(\rho)-m^*_{N^*}(\rho)=(1-C\rho/\rho_0)(m_N-m_N^*).
\label{eq:massdif}
\end{equation}

As for the $N^*$ width in the medium, we consider the two dominant decay
channels of the $N^*$ in the medium, namely $N^*\rightarrow\pi N$ and $NN^*\rightarrow\pi
NN$ in this calculation.
The other decay modes are shown to be negligible in our previous paper
\cite{ref:PRC66}. 
The $N^{*} \rightarrow \eta N$ channel does not contribute in the nuclear
medium due to the Pauli blocking effects on the decayed nucleon and
the $N^{*} \rightarrow \pi\pi N$ contribution is negligibly small in this model. 
The partial decay width for $N^*\rightarrow\pi N$
is calculated using the energy of the $N^*$ and obtained as,
\begin{equation}
  \Gamma_{\pi}(s)=3 {g^*_{\pi NN^*} \over 4\pi} {E_N + m_{N^*} \over
  \sqrt{s}} q \label{eq:Nstarwid},
\end{equation}
where $E_N$ and $q$ is the energy and the momentum of the final
nucleon on the mass shell in the rest frame of the $N^*$, respectively.
Similarly, we estimate the $NN^*\rightarrow\pi NN$ process within this
model following the formulation of Ref. \cite{ref:PRC44},
\begin{eqnarray}
  \lefteqn{\Gamma_{N^*N\rightarrow \pi NN} (s)=}    \\ \nonumber 
  &&  3 \beta^2 \left( {g_{\pi NN} \over 2
   m_N^*} \right)^2 \rho \int dp_1 p_1^3 \int {dp_2 \over (2 \pi)^3} 
   p_2  {m_N^* \over \omega_2}  \nonumber \\
  &&  \times
  {-\vec p_1^{\, 2} + 2m_N^* (\sqrt s - \omega_2 - m_N^*) \over 
   \left[ \left( {p_1^2 \over 2 m_N^*} \right)^2 - p_1^2 - m_\pi^2 \right]^2}
  \Phi(p_1,p_2) \ , \label{NNpidecay}
\end{eqnarray} 
where $p_1$ ($\omega_1$) and $p_2$ ($\omega_2$) 
are pion momenta (energies), $\Phi$ is the phase space variable defined
in \cite{ref:PRC44}. We define $\beta$ as,
\begin{equation}
\beta = { g_1  m_0 \over \langle \sigma \rangle m_N^* (m_{N^*}^* + m_N^*) }
   \chi ,
\end{equation}
with the effective coupling of $ \pi\pi NN^*$ through $\sigma$ meson 
in this model, which is $\chi \sim 1.29$.
This contribution is estimated to be typically fifteen MeV at the
saturation density, although it depends on the $\eta$ energy and $C$
parameter.  We include this channel in the present calculation.

We also mention the `naive' assignment case in the chiral doublet
model. The lagrangian density for the naive assignment is given by,
\begin{eqnarray}
{\cal L}&=& \sum_{j=1,2}\left[\bar{N}_ji\delslash N_j + a_j\bar{N}_j
(\sigma+i\gamma_5 \vec \tau \cdot \vec \pi)N_j\right] \nonumber\\
&+&a_3\left\{\bar{N}_2(\gamma_5 \sigma+i\vec\tau \cdot \vec\pi)N_1
-\bar{N}_1(\gamma_5\sigma+i\vec\tau\cdot\vec\pi)N_2\right\}
\label{eq:naiveL}
\end{eqnarray}
where $a_j(j=1,2,3)$ are the coupling constants.
In the physical base,
the masses of the $N$ and $N^*$ are given by
\begin{equation}
m^*_{N,N^*}=\frac{\mevsig}{2}(\sqrt{(a_1+a_2)^2 + 4a_3^2} \mp (a_1-a_2)).
\label{eq:mass_naive}
\end{equation}
The details of this model are discussed in Ref.  \cite{ref:PTP106}. As 
seen in
eq. (\ref{eq:mass_naive}), the mass difference of
$N$ and $N^*$ is expressed as a linear function of
$\langle\sigma\rangle$, which 
has exactly 
the same form as the mirror assignment case in Eq. (\ref{eq:massdif}). Therefore,
as shown later,
the general behavior of the $\eta$ optical 
potential obtained with
the naive model is similar to that of the mirror model,
since, essentially, 
the $N$-$N^*$ mass difference is responsible for the qualitative change of
the optical potential from attractive to repulsive in Eq. (\ref{eq:potential}).

As for the $N^*$ width in the medium, the  two decay channels, $N^*\rightarrow \pi N$
and $NN^*\rightarrow\pi NN$, are considered in the same manner as in the
mirror assignment case. 
In the naive assignment, the $\sigma NN^*$ coupling vanishes 
under a diagonalization of the mass matrix.
Hence, we consider additional terms given in Ref. \cite{ref:NPA640}, which
describe quadratic meson-baryon interaction including 
the $\pi\pi NN^*$ coupling to calculate the width.

\subsection{Chiral Unitary Model}

We explain briefly another type of chiral model
for baryon resonances,
the chiral unitary
approach \cite{ref:PLB550}, which is also used to describe the $N^*$
resonance in the nuclear medium and to obtain the $\eta$-nucleus
interaction.
In this approach, a
coupled channel Bethe-Salpeter equation for the meson-baryon scatterings 
is solved 
in vacuum, and the $N^*$ is generated dynamically as a resonance,
contrary to the chiral doublet model, where
the $N^*$ field appears in the lagrangian explicitly
as explained in the previous section.

To include the nuclear medium effects\cite{ref:PLB550}, they take into account
the Pauli blocking of the
intermediate nucleon states and use the in-medium propagators of the
intermediate mesons ($\pi,K,\eta$) and baryons ($N,\Lambda,\Sigma$).
The energy dependence of the each self-energy is treated in a
self-consistent manner.

In the present paper, we directly take the $\eta$ self-energy shown in
Fig.6 (c) of the second reference in Ref.  \cite{ref:PLB550}
calculated by the Valencia group
as the results of the chiral unitary approach, and use it to obtain the
$\eta$-nucleus optical potential.

\section{Numerical Results}
\label{sec:Result}
In this section, we show the numerical results on the $\eta$-nucleus
optical potential and the formation
cross sections of
$\eta$-nucleus systems using the different models for $\eta$-nucleus
interaction described in
section \ref{sec:model}.

\subsection{$\eta$-nucleus interaction}

\begin{figure}[hbt]
\epsfxsize=12cm
\centerline{
\epsfbox{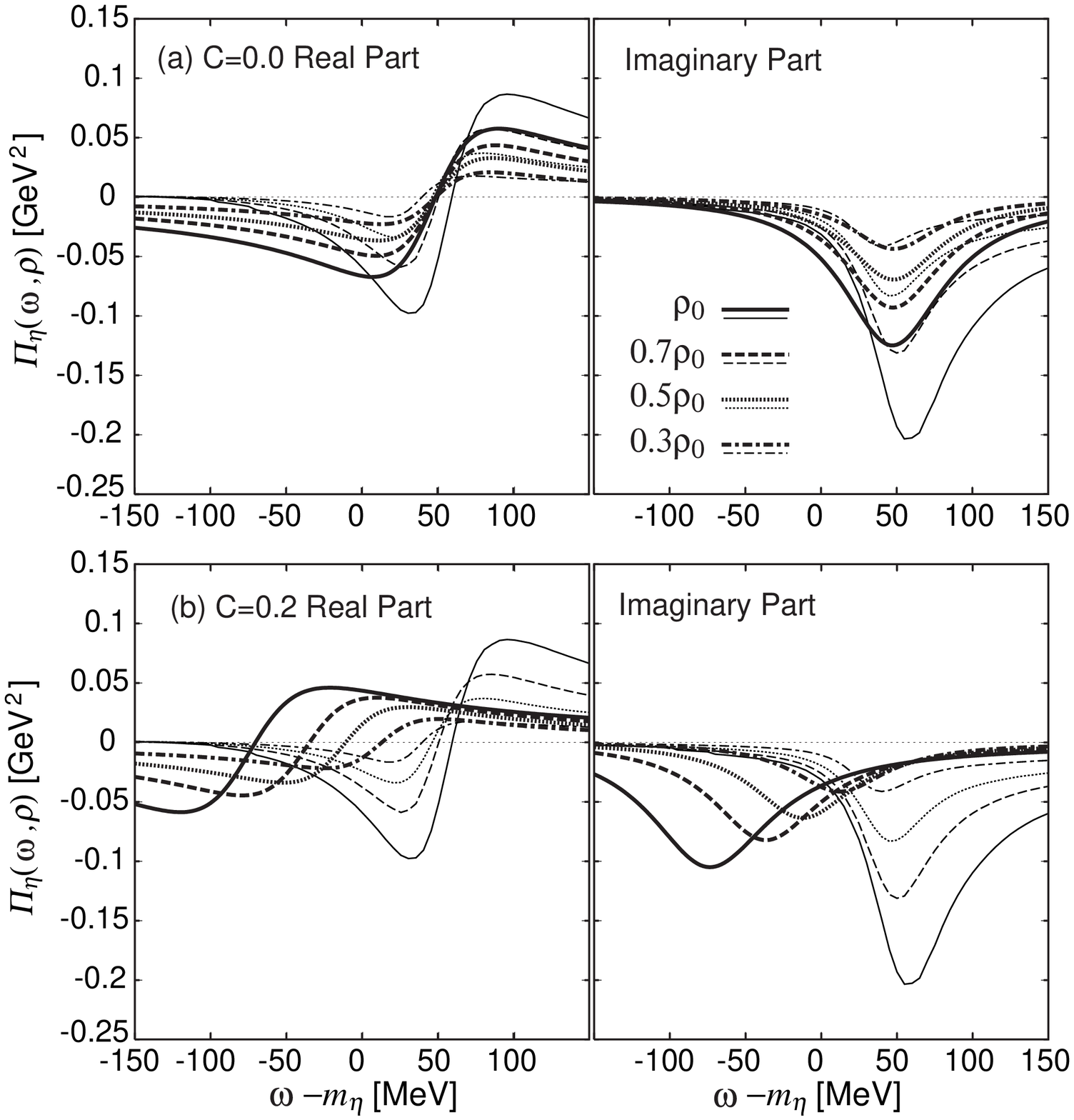}}
\caption{The $\eta$ self-energies are plotted as a function of
the $\eta$-energy for 4 nuclear density cases as indicated in the figure.
(a) The self-energies obtained by the chiral doublet model with the mirror
assignment for $C=0.0$ (thick lines) and those by the chiral unitary
approach (thin lines). (b) Same as (a) except for $C=0.2$ for the chiral
doublet model (thick lines).  The results with the chiral unitary approach are taken from
the
Fig.6 (c) of the
second reference in Ref.  \cite{ref:PLB550} and are the same for both (a) and (b).
\label{fig:self-energy}}
\end{figure}

The calculated $\eta$ self-energy in the nuclear medium depends on the
$\eta$ energy and the nuclear density in general.
We firstly show the calculated $\eta$
self-energies in different models
at finite nuclear density.
In Fig. \ref{fig:self-energy}, we show the $\eta$ self-energy at certain
nuclear densities as a function of the energy carried by the $\eta$.
We compare the self-energies obtained by the chiral doublet model
(the mirror assignment) with those obtained by the chiral unitary approach.
We show the results for $C=0.0$ case in Fig. \ref{fig:self-energy} (a)
and $C=0.2$ case in (b) for the chiral doublet model.
The results with the chiral unitary approach are directly taken from
Fig.6 (c) of the
second reference in Ref.  \cite{ref:PLB550} and are the same in both Figs.
\ref{fig:self-energy} (a) and (b).
In the case of $C=0.0$ in the chiral doublet model, since there is no
in-medium change of the sigma condensate $\langle\sigma\rangle$, the
properties of $N$ and $N^*$ do not change in the medium. Therefore this case
corresponds to the so-called '$t\rho$' approximation.
We find that the self-energies of the chiral doublet model with $C=0.0$
resemble those
of the chiral unitary approach.
On the other hand, the results with
$C=0.2$ show significant differences from the results of
the chiral unitary approach as can be seen in Fig.
\ref{fig:self-energy} (b). 
Both real and imaginary parts of the
self-energies of these two models show much
different energy dependence for all nuclear densities considered here.

\begin{figure}[hbt]
\epsfxsize=13cm
\centerline{
\epsfbox{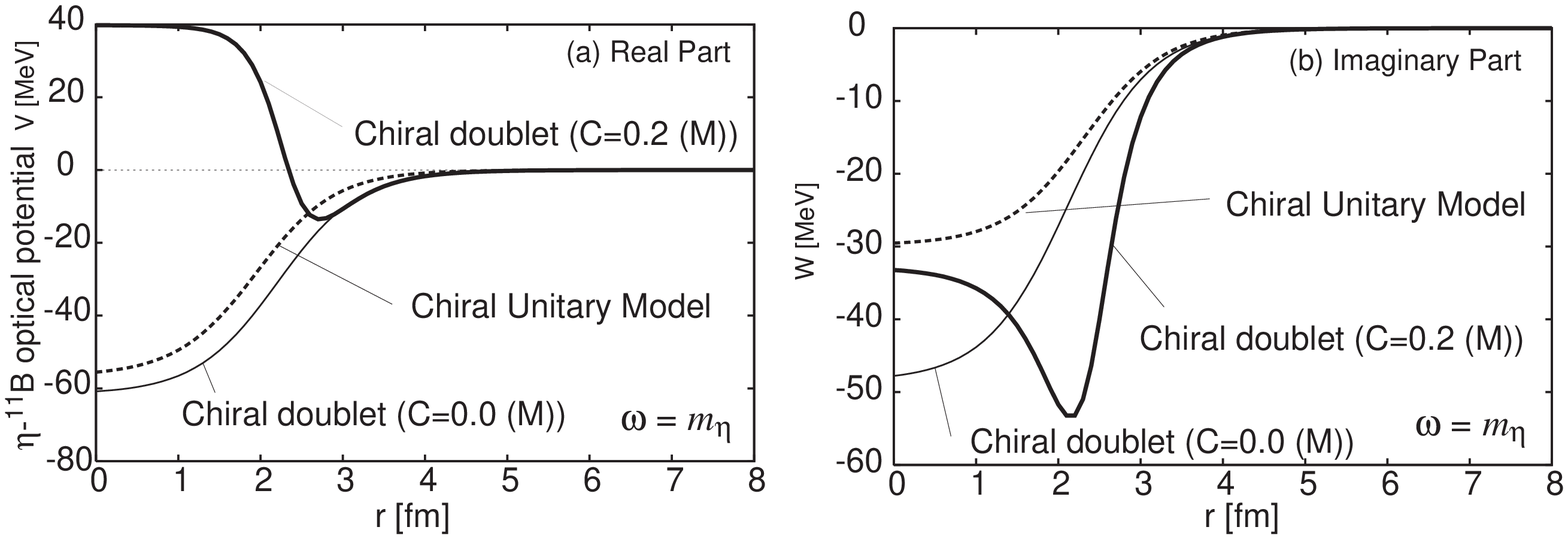}}
\caption{
The $\eta$-optical potentials for the $\eta$-$^{11}$B system as a function of
the radius
 coordinate $r$. The left and right figures show the real part and
 imaginary part, respectively. In both figures,
the potentials of the chiral doublet model with the mirror assignment are shown
 in the solid lines for $C=0.2$ (thick line) and $C=0.0$ (thin line).
The potential obtained using the chiral unitary model is shown in dotted
 line,
which is obtained from the results shown in Ref. \cite{ref:PLB550}
\label{fig:potential}}
\end{figure}

To see the consequences of these differences in the self-energy, we show
in Fig. \ref{fig:potential} the $\eta$-nucleus optical potential $U$
defined as;

\begin{equation}
U(r) = V(r) + i W(r) = \frac{1}{2 m_\eta} \Pi_\eta (m_\eta,\rho(r)),
\end{equation}

\noindent
where the $\eta$ energy is fixed to
be $m_\eta$. The nuclear density is assumed to be of an empirical
Woods-Saxon form here as;

\begin{equation}
\rho(r) = \frac{\rho_0}{1+{\rm exp} \left( \frac{r-R}{a} \right)},
\label{eq:ws}
\end{equation}

\noindent
where $R=1.18A^{1/3}-0.48$ (fm) and $a=0.5$ (fm) with the nuclear mass number
$A$.
We fix the $\eta$ energy to its mass $m_\eta$ here to see the $r$
dependence of the optical potential. In all other numerical results
shown in this paper, we use the appropriate energy dependent
$\eta$-Nucleus self-energies.

The optical potential is plotted in Fig. \ref{fig:potential} for
$\eta$-$^{11}$B system.  We find that the potential with $C=0.0$ resembles
that of the chiral unitary approach as expected from the behavior of
the self-energies, 
and that they are essentially attractive potential.  
However, the potential with $C=0.2$ has the repulsive
core inside the nucleus as reported in Ref.  \cite{ref:PRC66} and is much
different from the potential of the chiral unitary approach.

\subsection{$\eta$-mesic nucleus formation by the (d,$^3$He) reaction}

In this section, we evaluate the formation rate of the $\eta$-nucleus
system by the (d,$^3$He) reaction and show the calculated results for various
nuclear target cases.
In the (d,$^3$He) reaction spectroscopies, we only observe the emitted
$^3$He in the final state and obtain the double differential cross
section $d\sigma/d\Omega/dE$ as a function of the $^3$He energy.
The energy of $\eta\otimes$Nucleus system is evaluated from the $^3$He
kinetic energy and the properties of the $\eta$-Nucleus interaction can
be investigated from
the $d\sigma/d\Omega/dE$ data.
We use the Green function method to calculate the formation
cross sections of quasi-stable $\eta$-nucleus system \cite{ref:NPA435}.
All details of the application of the Green function method to the $\eta$
system formation are found in Refs. \cite{ref:PRC66,ref:EPJA6}. In this paper, we
consider T$_d$=3.5 GeV as the initial deuteron kinetic energy which
satisfies the recoilless condition for the $\eta$ production.

\begin{figure}[hbt]
\epsfysize=7cm
\centerline{
\epsfbox{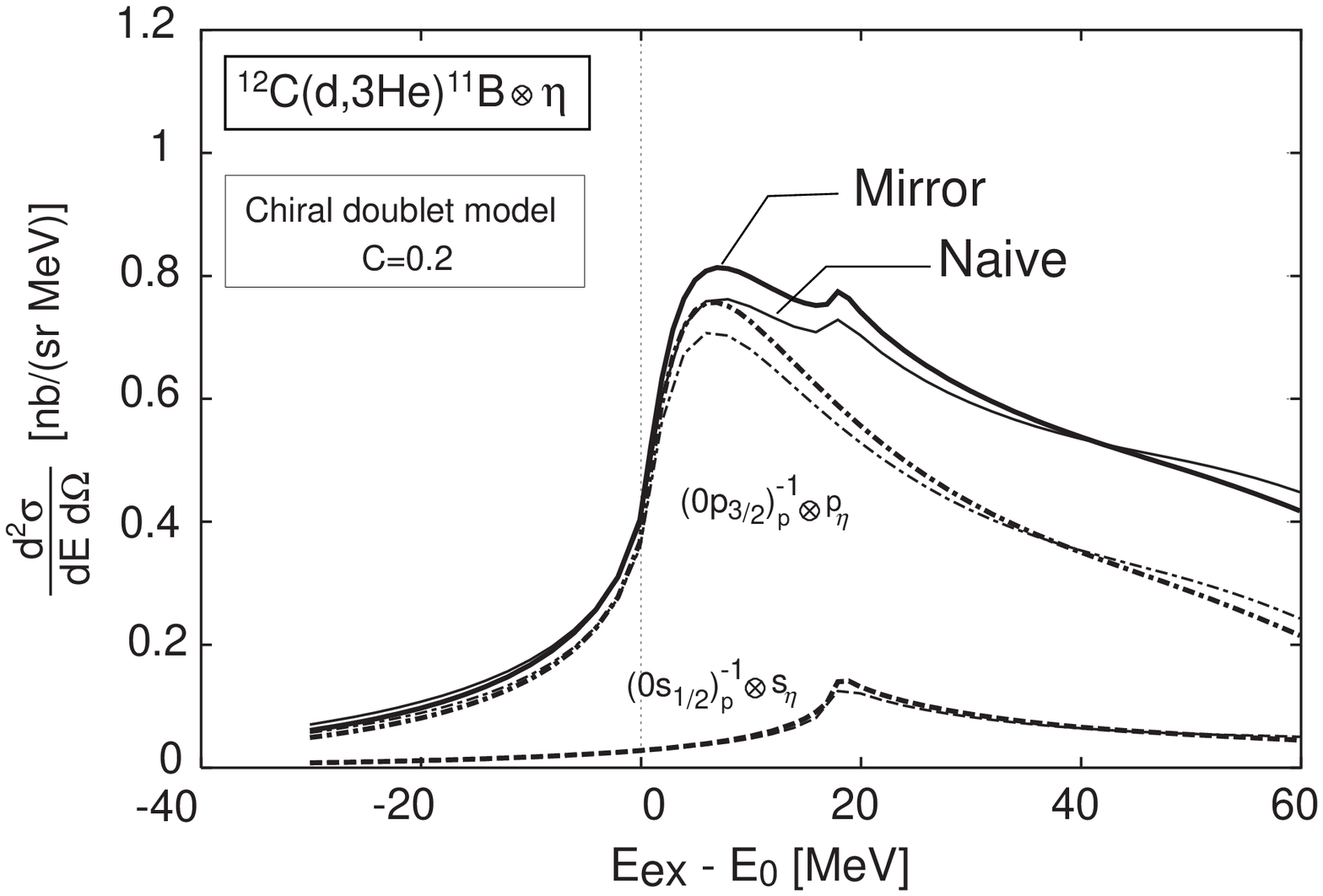}}
\caption{The calculated spectra of $^{12}$C(d,$^3$He)
reaction at T$_d$=3.5 GeV for the
formation of the $\eta$-$^{11}$B system are shown as functions of the excited energy
E$_{{\rm ex}}$ defined in the text. $E_0$ is the $\eta$ production
threshold energy. Thick lines show the results with the
mirror assignment and thin lines with the naive assignment. The dominant
 contributions 
from the $(0s_{1/2})^{-1}_p \otimes s_\eta$ and
the $(0p_{3/2})^{-1}_p \otimes p_\eta$ configurations are shown by
dashed lines and dash-dotted lines, respectively.
Here, the proton-hole states are indicated as $(n\ell_j)_p^{-1}$ and $\eta$
 states as $\ell_\eta$.
\label{fig:12C_target1}}
\end{figure}

We show the
$^{12}$C(d,$^3$He)$^{11}$B$\otimes \eta$ reaction cross sections for the
formation of the $\eta$-$^{11}$B system in the final state in
Fig. \ref{fig:12C_target1}. 
The spectra obtained are shown as functions of the
excited energy defined as;

\begin{equation}
E_{{\rm ex}}=m_\eta - B_\eta + (S_p(j_p)-S_p({\rm ground})),
\label{eq:E_ex}
\end{equation}

\noindent
where $B_\eta$ is the $\eta$ binding energy and $S_p(j_p)$ the proton
separation energy from the proton single particle level $j_p$.  The
$S_p({\rm ground})$ indicates the separation energy from the proton level
corresponding to the ground state of the daughter nucleus.
The nuclear density distributions are assumed to be the
empirical Woods-Saxon form defined in Eq. (\ref{eq:ws}).

Here, we briefly explain the general features of the (d,$^3$He) spectra
for the $\eta$ production using Fig.\ref{fig:12C_target1}. As shown
in the figure, the spectrum is dominated by two contributions which are
$(0s_{1/2})_p^{-1}\otimes s_\eta$ and $(0p_{3/2})_p^{-1}\otimes p_\eta$
configurations since the final states with the total spin $J\sim 0$ are
largely enhanced in the recoilless kinematics.
The $\eta$ production threshold with the $(0p_{3/2})_p^{-1}$
proton-hole state is indicate by the vertical dotted line at $E_{\rm
ex}-E_0=0$. The threshold with the $(0s_{1/2})_p^{-1}$ hole state, which
is the excited state of the daughter nucleus, is at $E_{\rm ex}-E_0=18$
MeV because of the difference of $S_p(j_p)$ in
Eq. (\ref{eq:E_ex}). Thus, the contributions from the bound $\eta$ states
appear in $E_{\rm ex}-E_0 < 0$ region with $(0p_{3/2})_p^{-1}$
state and in $E_{\rm ex}-E_0<18$ MeV region with $(0s_{1/2})_p^{-1}$
state.

In the present case of Fig.\ref{fig:12C_target1}, there are no bound
states and the strength in the 
bound region is due to the absorptive interaction of $\eta$-Nucleus system.
The existence of the imaginary part in the potential, which account for
the absorption of $\eta$ in nucleus, deform the shape of the
spectrum and provide
the certain strength in subthreshold region.
Hence, this subthreshold strength
has no relation to the existence of the bound states. If we have the
bound states with sufficiently narrow width, we will see the peak
structure in $\eta$ bound region, which is not seen in
Fig. \ref{fig:12C_target1}.

In the higher excitation energy region, the calculated spectra show the
contribution from the quasi-free (positive energy) $\eta$ production
with a proton-hole state. Since the recoilless condition is satisfied
only around $E_{\rm ex}-E_0\sim 0$, the cross section is smaller for the
higher excitation energy due to the larger momentum transfer even 
the emitted quasi-free $\eta$ has the larger phase space.

Hence, the peak structure shown in Fig.\ref{fig:12C_target1} is the
consequence of the reaction cross section around the $\eta$ production
threshold and does not have the direct connection to the existence of
the bound states. However, the whole shape of the spectrum reflects the
properties of the $\eta$-Nucleus interaction and provides important
information even if there are no quasi-stable bound peaks in the spectrum.

Going back to the discussion of our results, in
Fig.\ref{fig:12C_target1}, we show the 
calculated results by the chiral doublet model with both mirror and
naive assignments.
As expected in section \ref{subsec:CDM}, both
assignments predict the similar (d,$^3$He) spectra and show the
repulsive nature of the $\eta$-nucleus interaction.  Hereafter, we only
show the results 
with the mirror assignment since the both assignments provide
similar spectra.

\begin{figure}[hbt]
\epsfxsize=13cm
\centerline{
\epsfbox{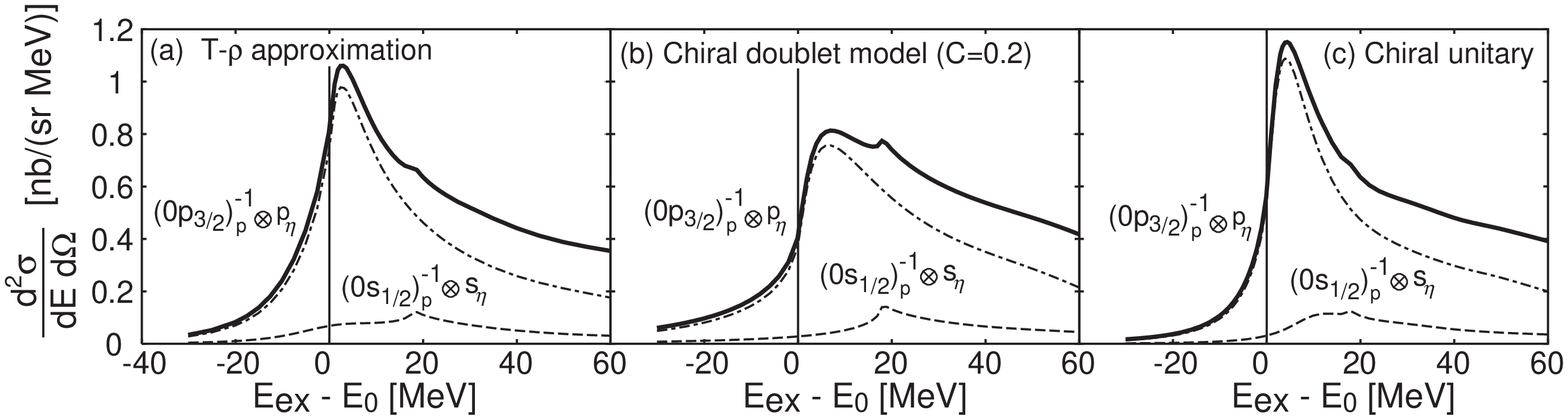}}
\caption{The calculated spectra of $^{12}$C(d,$^3$He)$^{11}$B$\otimes\eta$
reaction at T$_d$=3.5 GeV are shown as functions of the excited energy
E$_{{\rm ex}}$ defined in the text. $E_0$ is the $\eta$ production
threshold energy. The $\eta$-nucleus interaction is calculated by
(a) the $t \rho$ approximation, (b) the chiral doublet model with
$C=0.2$, and (c) the chiral unitary approach.  The total spectra are
shown by the thick solid lines, and
the dominant contributions
from the $(0s_{1/2})^{-1}_p \otimes s_\eta$ and
the $(0p_{3/2})^{-1}_p \otimes p_\eta$ configurations are shown by
dashed lines and dash-dotted lines, respectively.
Here, the proton-hole states are indicated as $(n\ell_j)_p^{-1}$ and $\eta$
 states as $\ell_\eta$.
\label{fig:12C_target2}}
\end{figure}

In Fig. \ref{fig:12C_target2},
we show the $^{12}$C(d,$^3$He)$^{11}$B$\otimes \eta$ spectra for three
different $\eta$-nucleus optical potentials.
In Fig. \ref{fig:12C_target2} (a), the spectra with the so-called $t \rho$
optical
potential, which are calculated by putting $C=0.0$ in the chiral
doublet model, are shown.  We show the spectra obtained by the chiral
doublet model
with $C=0.2$ in Fig. \ref{fig:12C_target2} (b). We can see in the figures
that the repulsive nature of the potential shifts the (d,$^3$He) spectrum
to the higher energy region compared to the $t \rho$ case.  In Fig.
\ref{fig:12C_target2} (c), we show the results by the chiral unitary
model.  As expected by the potential shape, the spectra with the chiral
unitary approach are shifted significantly to the lower energy region as in the $t \rho$
potential case as a result of the attractive potential.
We should mention here that we can see the contributions from the bound
$\eta$ states in Figs. \ref{fig:12C_target2}(a) and \ref{fig:12C_target2}(c)
as bumps in dashed lines around $E_{\rm ex}-E_0=10\sim 15$ MeV.
We have found that there exist certain discrepancies between the spectra
obtained with different chiral models, which are expected to be
distinguished by the experimental data.

Next we consider the case of an unstable nuclear target.
As shown in Fig. \ref{fig:potential}, since the chiral doublet model combines
the possible existence of the attractive $\eta$-nucleus interaction at
lower nuclear densities with the repulsive interaction at higher densities,
it will be interesting to study the $\eta$ mesic state in the unstable
nuclei with halo structure \cite{ref:JPG22}.
Here, we consider $^{11}$Li as an example of
the halo nuclei and evaluate the cross section of the $^{12}$Be(d,$^3$He)
reaction for the formation of the $\eta$-$^{11}$Li system in the final
state.

The density distribution of the $^{11}$Li is determined from the 
experimental interaction cross sections \cite{ref:PRL55}
by using the
cluster-orbital shell model approximation (COSMA)\cite{ref:COSMA,ref:PRC53}.
In this approximation,
the $^{11}$Li density is expressed as the sum of the $^9$Li-core and two valence
neutron densities.
For the $^9$Li-core, we use the Gaussian functional form for the 
proton and neutron densities,
which reproduce the experimental RMS radii as $R^p_{\rm rms}=R^n_{\rm
rms}=R^{\rm exp}_{\rm mat}$($^9$Li$)\ =\ 2.32\ $[fm] \cite{ref:PRL55}.
For halo density, we consider two possibilities for the orbital angular momentum
of the halo neutrons and apply the following two kinds of functional form
\cite{ref:COSMA};
\begin{eqnarray}
\rho^{\rm 1s}(r)=\frac{1}{\pi^{2/3}\alpha_{\rm 1s}^3}
\exp(-(r/\alpha_{\rm 1s})^2),\\
\rho^{\rm 1d}(r)=\frac{4r^4}{15\pi^{2/3}\alpha_{\rm 1d}^7}
\exp(-(r/\alpha_{\rm 1d})^2),
\end{eqnarray}
where the range parameter $\alpha$ is determined to be
$\alpha_{\rm 1s}=4.88$ [fm] and $\alpha_{\rm 1d}=3.2$ [fm]
\cite{ref:COSMA} so as to reproduce the experimental radius of
$^{11}$Li as
$R^{\rm exp}_{\rm mat}(^{11}{\rm Li})=3.2$ [fm] \cite{ref:PRL55}.

For the density distribution of the target nucleus $^{12}$Be, we sum up
the square of
the harmonic oscillator wave functions for all occupied states
to obtain the point nucleon density.
The oscillator parameter is determined by the experimental RMS radius of $^{12}$Be,
$R_{\rm rms}(^{12}{\rm Be})=2.59$ [fm] \cite{ref:PLB206}.
To get the charge (matter) distribution of the $^{12}$Be, we
fold the point nucleon density with the Gaussian nucleon density with
the size of $({R^{N}_{\rm rms}})^2=0.69$ [fm$^2$].


In Fig. 5, we show the calculated spectra for the formation of the
$\eta \otimes ^{11}$Li system by the $^{12}$Be(d,$^3$He) reaction.
Here, we assume that the halo neutrons are in the $1s$ state
and use the Eq. (17) for the halo neutron density of the COSMA.
We compare the results of the chiral doublet model with those of the chiral
unitary model.  In this case, we can see the repulsive nature of the chiral
doublet
prediction and the differences of the results of these models again.
Both results have qualitatively the same features as the results shown in
Fig. 4 for the $^{12}$C target cases.  However, the contributions from the
$(0s_{1/2})^{-1}_p \otimes s_\eta$ configurations are relatively enhanced in
the $^{12}$Be target case because of the difference of the single particle
proton
configuration in the target. In the total spectrum, we can clearly see the
cusp due to the
$(0s_{1/2})^{-1}_p \otimes s_\eta$ configurations, which could be
interesting and useful to
identify the $\eta$ contributions in experiments.

To see the low density halo contributions clearly, we show the results with
halo-neutrons in $1s$ state (Eq. (17)), in $1d$ state (Eq. (18)), and
without halo neutron
cases in Fig. 6 for the chiral doublet model and the chiral unitary model.
In both models,
we cannot see significant differences in the spectra due to the
difference of
the halo neutron states, $1s$ or $1d$, in dominant contributions as shown in
the figure.
The spectra calculated without halo neutrons, which are thought to
be equivalent to the contribution from the core $^9$Li, have slightly larger
cross sections
in all cases considered here.  We think this enhancement is due to the lack
of the distortion effects
in the final states from the halo neutrons.

From the results shown in Figs. 5 and 6, we think that it is difficult to
observe the
characteristic spectra due to the
existence of the wide low density region with the realistic
halo neutron density distribution of $^{11}$Li.  In order to see the
low-density-attractive nature of the optical potential predicted by the
chiral doublet model,
we need to consider the nuclei with more `effective' halo than $^{11}$Li,
which has larger spatial dimension and includes more neutrons, and may exist
in the
heavier mass region \cite{Meng98}.

In all results shown above, we consider the target nuclei that include
protons both in
$s_{1/2}$ and $p_{3/2}$ states.  In the (d,$^3$He) reactions with the
recoilless condition, the substitutional configurations are known to be
largely populated.   Hence, to consider the
target nucleus which includes only $s_{1/2}$ protons is interesting because
the spectrum will
be dominated by the only contribution $(0s_{1/2})^{-1}_p \otimes s_\eta$.
We can expect to
deduce the information of the $\eta$-nucleus interaction very easily without
decomposing
the spectrum into subcomponents.  For this purpose, we consider the $^4$He
as a target nucleus and
calculate the (d,$^3$He) spectrum for the formation of the $\eta$-$^3$H
system in the final state.
Of course, we are aware of the lack of the few-body treatment in our
formalism and we should
improve it for a more quantitative calculation.
However, we think it is still extremely interesting to evaluate
the spectrum for the
$^4$He target case to see the advantages to observe the spectra with the
single dominant subcomponents.

We show the calculated results of the  $^{4}$He(d,$^3$He)$^{3}$H$\otimes \eta$
reaction in Fig. 7.  As we expected, the results are completely
dominated by the single components and will be easily related to the
behavior of the
$\eta$-nucleus interaction and $N^*$ properties in the medium.
We can see the clear differences of the three cases here.

Finally, we show the results with the heavy target $^{40}$Ca case
in Fig. 8.  In the heavier targets, we have larger possibilities to have bound
states and
the larger medium effects.  However, it seems to be very difficult to deduce
the
clear information from the spectra since many configurations
$(n\ell_{j})^{-1}_p \otimes \ell_\eta$ contribute to the spectra and they can not
be distinguished
because of the large width.  We calculate the
$^{40}$Ca(d,$^3$He)$^{39}$K$\otimes \eta$
spectrum with the chiral doublet model and see the experimental
feasibilities to use the
heavy target.
We find that the whole spectrum is shifted according to the change
of the $C$ parameter in the model and we may be able to deduce the
average  strength of the potential from the position of the whole
spectrum. 
However, as we expected, it seems extremely difficult and almost
impossible to
distinguish the each contribution from the full spectrum.

\section{Conclusion}

We have studied the properties of $\eta$-nucleus interaction
and their experimental implications.  We obtain the
$\eta$-nucleus optical
potential by postulating the $N^*(1535)$ dominance in the $\eta$-$N$ system.
The $N^*(1535)$
properties in the nuclear medium are taken into account by two kinds of the
chiral effective models,
the chiral doublet model and the chiral unitary model.

We find that the two kinds of chiral models lead to
the different properties of $\eta$-nucleus interaction
as a result of the qualitatively
different properties of the
$N^*$ in nuclear medium.
Hence, the studies of the $\eta$-nucleus interaction can be
connected to the properties of the $N^*$ in the medium
and the information of the in-medium baryon chiral symmetries.  Especially,
we should stress here that the chiral doublet model lead to a unique
shape of the $\eta$-nucleus optical potential which change its nature from
attractive to repulsive for higher nuclear densities.  It could be
extremely
interesting to confirm the existence (or non-existence) of this curious
shape of the potential experimentally.

To investigate the experimental feasibility, we have calculated the (d,$^3$He)
spectra for the formation
of the $\eta$-nucleus systems in the final states. This (d,$^3$He) spectroscopy is
an established experimental method in the studies of the pionic bound
systems.  We have studied theoretically the (d,$^3$He) spectra in a
comprehensive manner and 
concluded that
we can deduce the new information of $\eta$-nucleus interaction from the
(d,$^3$He) experiment,
and by knowing the nature of the $\eta$-nucleus 
optical potential,
we will be able to study the in-medium 
properties of the $N^*$.
We believe that this research helps much the experimental activities for
the studies of the
$\eta$-nucleus systems, and the understanding of the baryon chiral
symmetries and its medium modifications.


\begin{figure}[hbt]
\epsfysize=7cm
\centerline{
\epsfbox{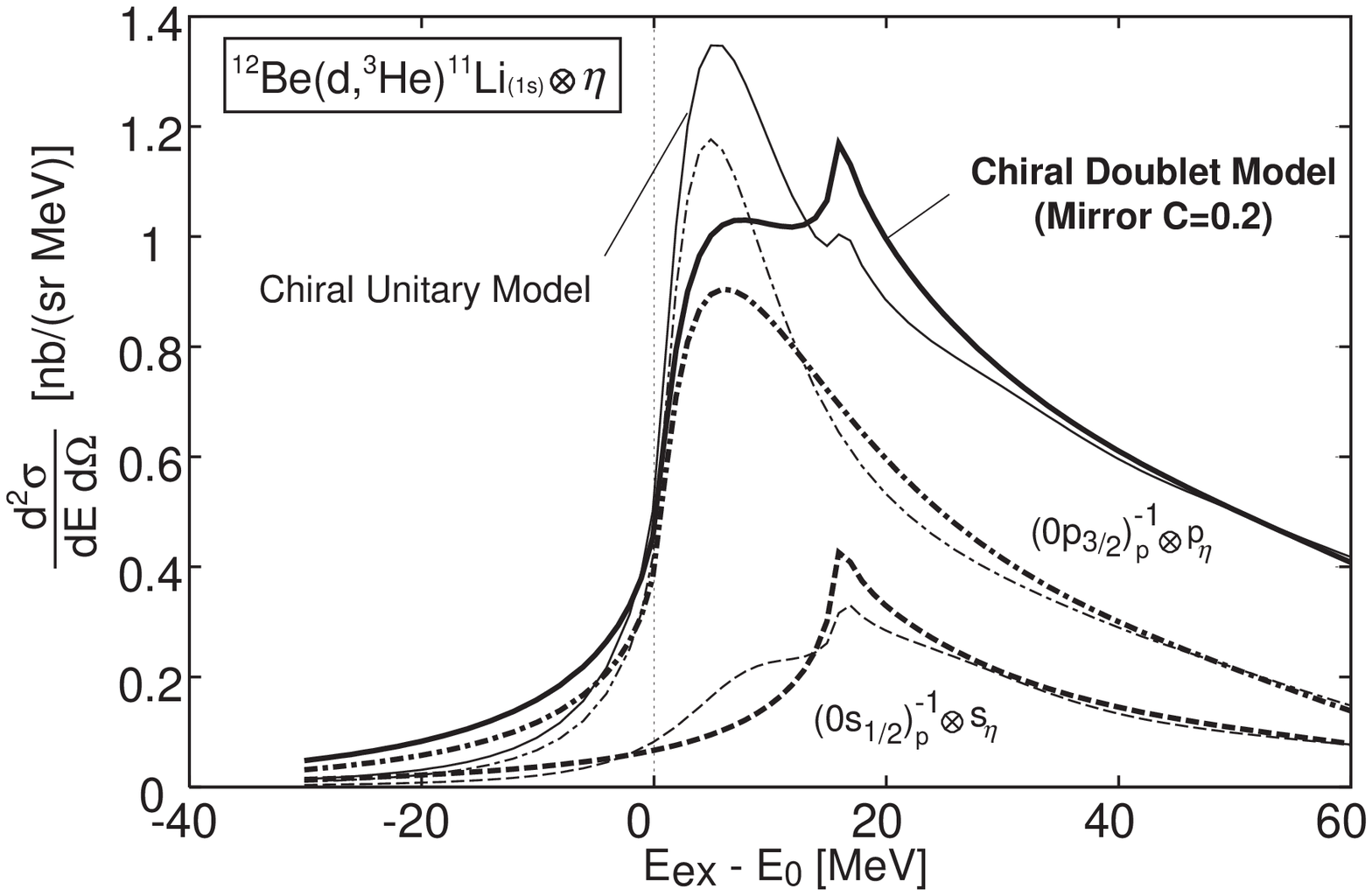}}
\caption{The calculated spectra of $^{12}$Be(d,$^3$He)$^{11}$Li$\otimes \eta$ reaction
at T$_d$=3.5GeV are shown
as functions of the excited energy E$_{\rm ex}$ defined in the text.  E$_0$
is the $\eta$ production threshold energy.
The $\eta$-nucleus interaction is calculated by the chiral doublet model
with the mirror assignment
with the parameter $C=0.2$ (thick lines), and the chiral unitary model (thin
lines).
The total spectra are shown by the solid lines, and the contributions from
the $(0s_{1/2})^{-1}_p \otimes s_\eta$ and the $(0p_{3/2})^{-1}_p \otimes
p_\eta$ configurations are shown by dashed lines and dashed-dotted lines,
respectively.
Here, the proton-hole states are indicated as $(n\ell_j)_p^{-1}$ and $\eta$
 states as $\ell_\eta$.
The $\rho ^{1s}$ form is used as the halo neutron density in the unstable
$^{11}$Li distribution (see text).
\label{fig:12Be_target1}}
\end{figure}

\begin{figure}[hbt]
\epsfxsize=13cm
\centerline{
\epsfbox{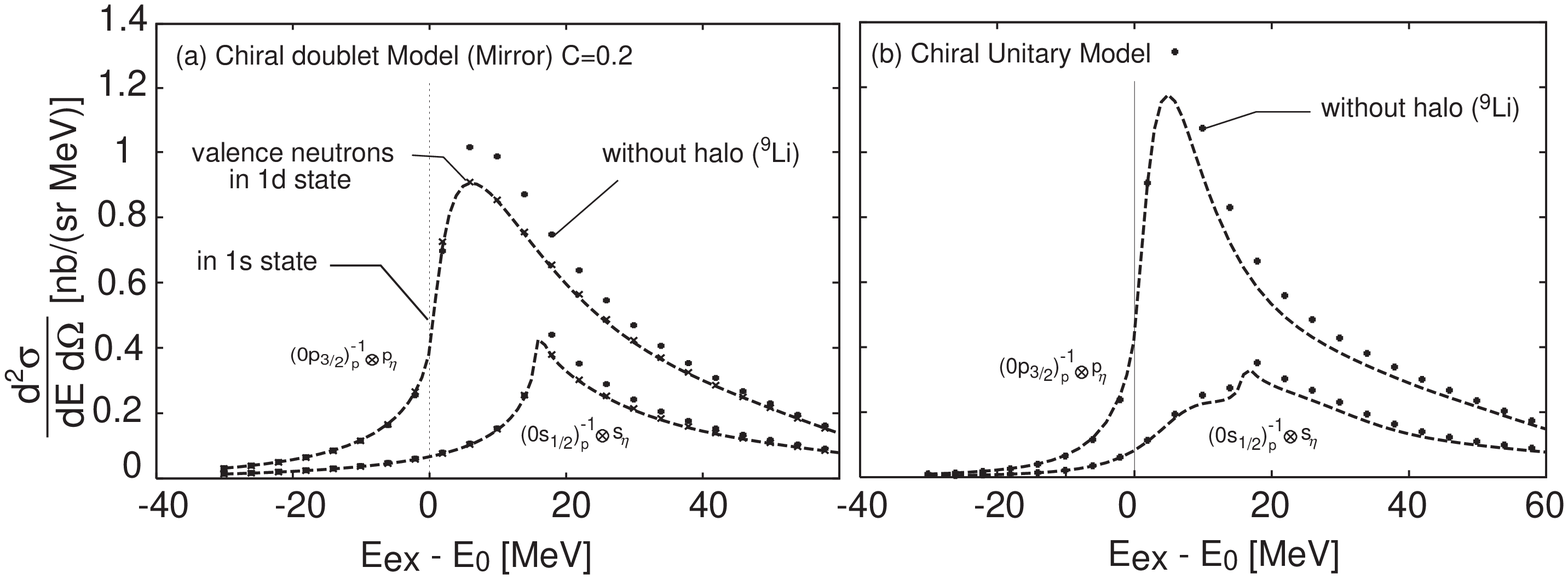}}
\caption{The calculated spectra of $^{12}$Be(d,$^3$He)$^{11}$Li$\otimes \eta$ reaction
at T$_d$=3.5GeV are shown
as functions of the excited energy E$_{\rm ex}$ defined in the text.  E$_0$
is the $\eta$ production threshold energy.
The $\eta$-nucleus interaction is calculated by (a) the chiral doublet model
with the mirror assignment with the parameter $C=0.2$, (b) and the chiral
unitary model.
In each figure, the contributions from the $(0s_{1/2})^{-1}_p \otimes
s_\eta$ and the $(0p_{3/2})^{-1}_p \otimes p_\eta$ configurations are
 shown.
Here, the proton-hole states are indicated as $(n\ell_j)_p^{-1}$ and $\eta$
 states as $\ell_\eta$.
The dashed lines indicate the results with the $\rho ^{1s}$
and the crosses with the $\rho ^{1d}$ halo neutron density distributions.
The dots indicate the
results from $^9$Li core calculated without halo neutron density.
\label{fig:12Be_target2}}
\end{figure}
%
%
%
%
\begin{figure}[hbt]
\epsfysize=7cm
\centerline{
\epsfbox{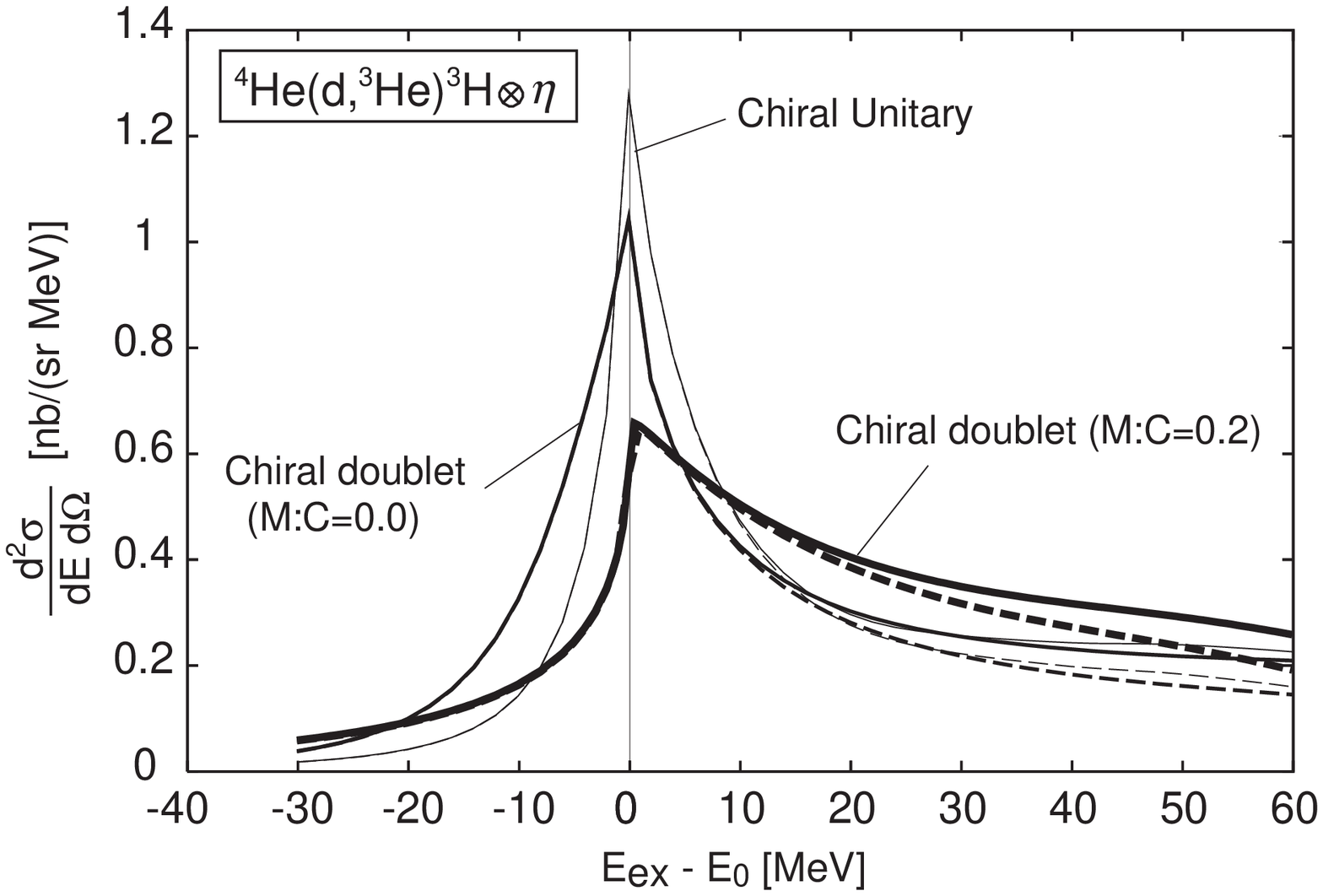}}
\caption{The calculated spectra of $^{4}$He(d,$^3$He)$^{3}$H$\otimes \eta$ reaction
at T$_d$=3.5GeV are shown
as functions of the excited energy E$_{\rm ex}$ defined in the text.  E$_0$
is the $\eta$ production threshold energy.
The $\eta$-nucleus interaction is calculated by the chiral doublet model
with the mirror assignment with the parameter $C=0.2$ (thick lines) and
$C=0.0$ (medium lines),
and the chiral
unitary model (thin lines).
In each figure, the contribution from the $(0s_{1/2})^{-1}_p \otimes
s_\eta$ configuration is shown as the dashed line.
Here, the proton-hole states are indicated as $(n\ell_j)_p^{-1}$ and $\eta$
 states as $\ell_\eta$.
\label{fig:4He_target}}
\end{figure}
%
%
%
\begin{figure}[hbt]
\epsfxsize=13cm
\centerline{
\epsfbox{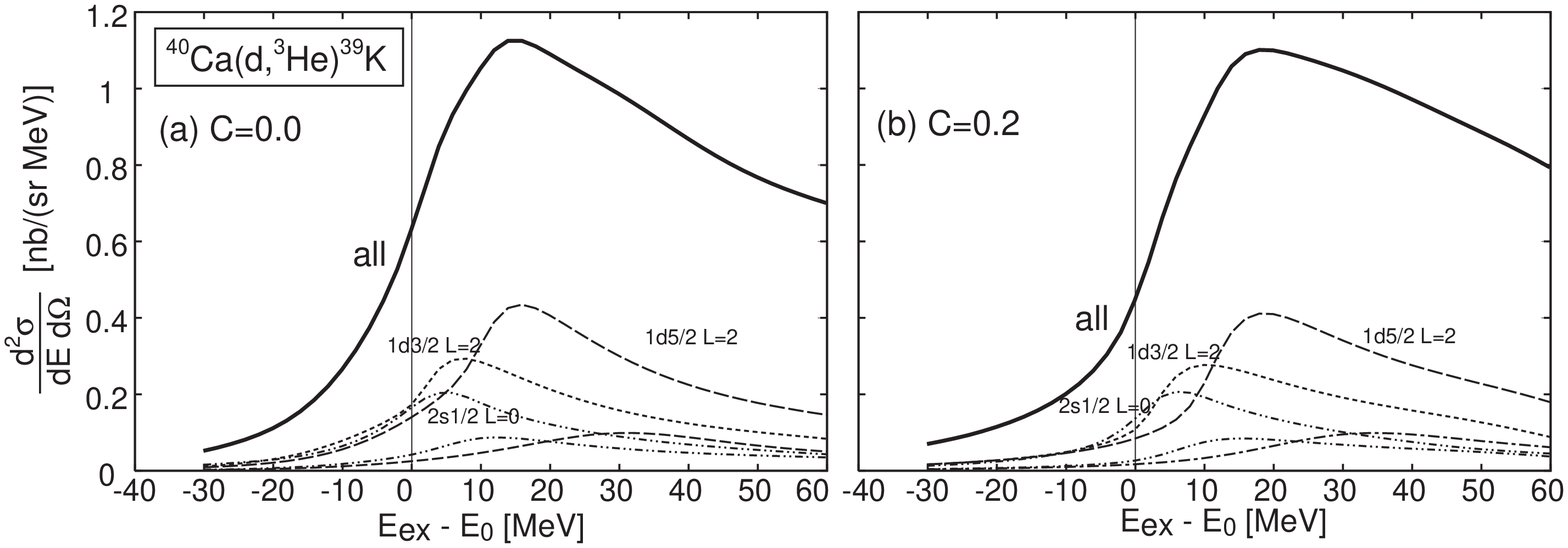}}
\caption{The calculated spectra of $^{40}$Ca(d,$^3$He)$^{39}$K$\otimes \eta$ reaction
at T$_d$=3.5GeV are shown
as functions of the excited energy E$_{\rm ex}$ defined in the text.  E$_0$
is the $\eta$ production threshold energy.
The $\eta$-nucleus interaction is calculated by the chiral doublet model
with the mirror assignment with the parameter (a) $C=0.0$ and (b) $C=0.2$.
In each figure, the full spectrum is shown by the thick solid line and
dominant sub-components
are also shown as indicated in the figure.
\label{fig:40Ca_target}}
\end{figure}
%
%


\section{Acknowledgements}
This work is partly supported by Grands-in-Aid for Scientific Research
(C) of the Japan Society for the Promotion of Science (JSPS), No. 14540268.


\begin{thebibliography}{99}
\bibitem{ref:PR287}
For example; C.J. Batty, E. Friedman and A. Gal, Phys. Rep. {\bf 287}, 385
(1997).

\bibitem{ref:PR247}
See the reviews; T. Hatsuda and T. Kunihiro, Phys. Rep. {\bf 247}, 221 (1994);
 G. E. Brown and M. Rho, $ibid$ {\bf 269}, 333 (1996).

\bibitem{ref:PLB514}
P. Kienle and T. Yamazaki, Phys. Lett. {\bf B514}, 1(2001);
H. Geissel et al., Phys. Lett. {\bf B549}, 64 (2002);
K. Suzuki et al., nucl-ex/0211023.

\bibitem{ref:APP31}
W. Weise, Acta Physica Polonica {\bf 31}, 2715 (2000);
E. E. Kolomeitsev, N. Kaiser and W. Weise, nucl-th/0207090

\bibitem{ref:PRC61}
S. Hirenzaki, Y. Okumura, H. Toki, E. Oset and A. Ramos, Phys.Rev. {\bf C61},
055205 (2000)

\bibitem{ref:PRC66}
D. Jido, H. Nagahiro and S. Hirenzaki, Phys. Rev. {\bf C66}, 045202
 (2002);
Proc. of PANIC02, Osaka, Japan (2002), Nucl. Phys. A in press,
nucl-th/0211085.

\bibitem{ref:PLB550}
C. Garcia-Recio, J. Nieves, T. Inoue and E. Oset, Phys. Lett. {\bf B550}, 47-54
(2002);
T. Inoue and E. Oset, Nucl. Phys. {\bf A710}, 354-370 (2002).

\bibitem{ref:NPA650}
F. Klingl, T. Waas and W. Weise, Nucl. Phys. {\bf A650}, 299 (1999).

\bibitem{ref:NPA1991}
H. Toki, S. Hirenzaki and T. Yamazaki, Nucl. Phys. {\bf A530}, 679 (1991);
S. Hirenzaki, H. Toki and T. Yamazaki, Phys. Rev. {\bf C44}, 2472 (1991).

\bibitem{ref:PL1988}H. Toki and T. Yamazaki, Phys. Lett. {\bf B213}, 129(1988);
H. Toki, S. Hirenzaki, T. Yamazaki and R. S. Hayano,
 Nucl. Phys. {\bf A501}, 653. (1989).

\bibitem{ref:K.Itahashi}
H. Gilg {\it et al.}, Phys. Rev. {\bf C62}, 025201 (2000);
K. Itahashi {\it et al.}, Phys. Rev. {\bf C62}, 025202 (2000).

\bibitem{ref:Geiseel}
H. Geiseel {\it et al.}, Phys. Rev. Lett. {\bf 88}, 122301 (2002).

\bibitem{ref:EPJA6}
R. S. Hayano, S. Hirenzaki and A. Gillitzer, Eur. Phys. J. {\bf A6}, 99
 (1999).

\bibitem{ref:PLB443}
K. Tsushima, D. H. Lu, A. W. Thomas and K. Saito, Phys. Lett. {\bf B443}, 26
 (1998)

\bibitem{ref:GSI}
R. S. Hayano, A. Gillitzer {\it et al.}, GSI proposal S214(1997).

\bibitem{ref:PRD66}
Particle data group, Phys. Rev. {\bf D66}, 01001 (2002).

\bibitem{ref:PRD39}
C. DeTar and T. Kunihiro, Phys. Rev.  {\bf D39}, 2805 (1989).

\bibitem{ref:PTP106}
D. Jido, M. Oka and A. Hosaka, Prog. Theor. Phys. {\bf 106}, 873 (2001);
D. Jido, Y. Nemoto, M. Oka and A.Hosaka, Nucl. Phys. {\bf A671}, 471
 (2000).

\bibitem{ref:PLB224}
T. Hatsuda and M. Prakash, Phys. Lett. {\bf B224}, 11 (1989)

\bibitem{ref:NPA640}
H. Kim, D. Jido and M. Oka, Nucl. Phys. {\bf A640}, 77 (1998).

\bibitem{ref:PRC44}
H. C. Chiang, E. Oset, and L. C. Liu, Phys. Rev. {\bf C44}, 738 (1991).

\bibitem{Nemoto:1998um}
Y.~Nemoto, D.~Jido, M.~Oka and A.~Hosaka,
Phys.\ Rev.\ D {\bf 57}, 4124 (1998).

\bibitem{Jido:1999hd}
D.~Jido, T.~Hatsuda and T.~Kunihiro,
Phys.\ Rev.\ Lett.\  {\bf 84}, 3252 (2000).

\bibitem{ref:PRL82}
T. Hatsuda, T. Kunihiro, and H. Shimizu, Phys. Rev. Lett. {\bf 82}, 2840
 (1999); D. Jido, T. Hatsuda, and T. Kunihiro, Phys. Rev. {\bf D63},
 011901 (2000).

\bibitem{ref:NPA435}
O. Morimatsu and K. Yazaki, Nucl. Phys. {\bf A435}, 727 (1985), Nucl. Phys
{\bf A483}, 493 (1988).

\bibitem{ref:JPG22}
For example, I. Tanihata, J. Phys. {\bf G22}, 157 (1996).

\bibitem{ref:PRL55}
I. Tanihata {\it et al.}, Phys. Rev. Lett. {\bf 55}, 2676(1985).

\bibitem{ref:COSMA}
M. V. Zhukov {\it et al.}, Phys. Rep. {\bf 231}, 151(1993).

\bibitem{ref:PRC53}
A. A. Korsheninnikov {\it et al.}, Phys. Rev. {\bf C53}, R537(1996).

\bibitem{ref:PLB206}
I. Tanitaha, {\it et al.}, Phys. Lett. {\bf B206}, 592(1988).

\bibitem{Meng98}
J. Meng and P. Ring, Phys. Rev. Lett. 80, 460 (1998).

\end{thebibliography}
\end{document}